\documentclass[11pt]{article}

\newcommand{\plb}[3]{Phys. Lett. {\bf B#1} (#2) #3} 
\newcommand{\prl}[3]{Phys. Rev. Lett. {\bf #1} (#2) #3}
\newcommand{\prd}[3]{Phys. Rev. {\bf D#1} (#2) #3}
\newcommand{\npb}[3]{Nucl. Phys. {\bf B#1} (#2) #3}
\newcommand{\npbps}[3]{Nucl. Phys. {\bf B}(Proc. Suppl.) {\bf #1} (#2) #3}

\title{
\hfill
\parbox{4cm}{\normalsize KUNS-1526\\ HE(TH)~98/12\\
{\tt  hep-lat/9808026}}\\
\vspace{1cm}
Axial vector current of exact chiral symmetry \\ on the lattice 
\author{
Yoshio Kikukawa\thanks{e-mail address:
kikukawa@gauge.scphys.kyoto-u.ac.jp}
\\
{\normalsize\em Department of Physics, Kyoto University 
}\\
{\normalsize\em Kyoto 606-01, Japan}
\\
\\
and 
\\
\\
Atsushi Yamada\thanks{e-mail address:
atsushi@hep-th.phys.s.u-tokyo.ac.jp}
\\
{\normalsize\em Department of Physics, University of Tokyo
}\\
{\normalsize\em Tokyo 113, Japan}
}
\date{\normalsize August, 1998}
}

\begin{document}
\maketitle

\begin{abstract}
We discuss the exact chiral symmetry and its spontaneous breakdown 
in lattice QCD 
with the Dirac operators satisfying the Ginsparg-Wilson relation. 
The axial vector current, which turns out to be related to the vector 
current simply by the insertion of the operator $\gamma_5 (1-aD)$,
is explicitly constructed in the case of the Neuberger-Dirac operator. 
We also consider an Euclidean proof of the Nambu-Goldstone theorem 
using the Ward-Takahashi identity of this symmetry.
\end{abstract}

\newpage

Recently Neuberger has proposed a Dirac operator which 
describes exactly massless fermions 
on a lattice \cite{overlap-Dirac-operator,GW-overlap-D} 
based on the overlap formalism of chiral 
determinant \cite{overlap, odd-dim-overlap}. 
Its explicit form is known as 
\begin{equation}
a  D_{nm}= \left( 1 + \gamma_5 \frac{H}{\sqrt{H^2}} \right)_{nm} ,
\end{equation}
where $H$ is the hermitian Wilson-Dirac operator defined by
\begin{equation}
H = \gamma_5 
\left\{ 
 \frac{1}{2}\gamma_\mu \left( \nabla_\mu -\nabla_\mu^\dagger \right)
+\frac{a}{2}\nabla_\mu\nabla_\mu^\dagger - \frac{1}{a} m_0 
\right\}, \quad ( 0 < m_0 < 2) .
\end{equation}
Here $SU(N)$ gauge group is assumed and 
$\nabla_\mu$ is the gauge covariant forward difference operator.

This Dirac operator satisfies the Ginsparg-Wilson 
relation \cite{ginsparg-wilson-rel,GW-fixed-point-D,GW-overlap-D,
almost-massless}, 
\footnote{Another Dirac operator 
which satisfies the Ginsparg-Wilson relation has been 
proposed by Hasenfratz et. al. \cite{GW-fixed-point-D,
fixed-point-Dirac-operator-I, 
fixed-point-Dirac-operator-II, 
index-finite-lattice,no-tuning-mixing}
based on the renormalization group method 
\cite{fixed-point-action-YM,fixed-point-action-fermion}.
As to this fixed point Dirac operator, the authors refer the reader
to the above cited references. }
\begin{equation}
  D_{nm} \gamma_5 + \gamma_5 D_{nm} = \sum_l a D_{nl} \gamma_5 D_{lm} .
\end{equation}
This relation guarantees that the effects of the chiral symmetry 
breaking terms in the Dirac operator appear only in local terms 
and physically irrelevant. 
This is the clue to escape the Nielsen-Ninomiya 
theorem \cite{nielsen-ninomiya}.

Subsequently, L\"uscher pointed out that the Ginsparg-Wilson 
relation implies an exact symmetry of the fermion 
action \cite{exact-chiral-symmetry}. The chiral transformation
proposed is given as 
\begin{equation}
\label{eq:exact-chiral-symmetry}
\delta \psi_n = \sum_m \gamma_5 
    \left( 1 - \frac{a}{2} D \right)_{nm} \psi_m , 
\quad
\delta \bar \psi_n = \sum_m \bar \psi_m 
 \left( 1 - \frac{a}{2} D\right)_{mn} \gamma_5 .
\end{equation}
This chiral transformation can be extended to the flavor non-singlet 
chiral transformations. L\"uscher also observed that 
for the flavor-singlet chiral transformation the 
functional integral measure is not invariant 
and its anomalous variant is indeed given in terms of the index 
of the Dirac operator.


Locality of the Neuberger-Dirac operator in the presence of gauge field 
is not quite obvious, although it provides a completely satisfactory 
local solution of the Ginsparg-Wilson relation 
in free theory \cite{almost-massless,exact-chiral-symmetry}. 
This question has been examined analytically and numerically by 
several authors. 
Hernandes, Jansen and L\"uscher have given a proof of the locality 
for a certain set of bounded small gauge fields and also for the case 
with an isolated zero mode of $H$ 
\cite{locality-overlap-D-small-su3}. 
Locality in dynamically generated gauge fields has been examined 
by same authors for $SU(3)$ and by Montvay for 
$SU(2)$ \cite{locality-overlap-D-su2}.
According to the former investigation, the Neuberger-Dirac operator 
seems local even with the gauge fields for $\beta \ge 6.0$.

The question of the practical use of the Neuberger-Dirac operators
also does not seem to be serious, 
although the Dirac operator involves the inverse square root of $H^2$ 
and appears at first sight to require the full storage of the size of 
the Dirac operator. In this respect, an approximate formula has been 
given by Neuberger \cite{practical-use-overlap-D}, 
for which the conjugate gradient method for
the (hermitian) Wilson-Dirac operator is applicable 
with the help of the shift technique for different 
fermion masses \cite{one-stroke-technique}. 
Other approaches have also been 
investigated 
by Chiu \cite{Newton-method-for-overlap-D} and
by Edwards, Heller and Narayanan \cite{practical-use-overlap-D-4dim}.

The weak coupling perturbation theory of the massless QCD 
with the Neuberger-Dirac operator has been discussed 
by the authors \cite{weak-coupling-expansion-overlap-D}.  
The axial $U(1)$ anomaly is evaluated explicitly and is shown 
to have the correct form of the topological charge density 
for perturbative backgrounds, supplementing the original 
calculation in \cite{ginsparg-wilson-rel}.

In this paper, we discuss the axial Ward-Takahashi identity 
associated with the exact chiral symmetry in some detail. 
We first consider the axial vector current which, 
following the Noether's procedure, is defined through the variation of 
the action by the ``local'' chiral transformation, 
\begin{eqnarray}
\label{eq:axial-current-divergence}
\delta_5 S_F 
&=& a^4 \sum_{nml}
 \bar \psi_n \left\{ 
 D_{nl} \alpha_l 
 \gamma_5 \left( 1 - \frac{a}{2} D \right)_{lm} 
+\left( 1 - \frac{a}{2} D\right)_{nl} \gamma_5  
\alpha_l D_{lm} \right\} \psi_m  \nonumber\\
&=& a^4 \sum_{nml}
 \bar \psi_n \left\{ 
D_{nl} \alpha_l - \alpha_n D_{nl}
\right\} \gamma_5 \left(1-a D\right)_{lm} \psi_m 
\nonumber\\
&=& a^4 \sum_l \left\{ \partial_\mu \alpha_l 
 \sum_{nm} \bar \psi_n K^5_{l\mu}(n,m) \, \psi_m \right\}.
\end{eqnarray}
Here we have used the Ginsparg-Wilson relation 
to derive the second line.
As we see from this structure of the variation of the action, 
the axial vector current may be constructed following the general 
procedure discussed by Ginsparg and 
Wilson \cite{ginsparg-wilson-rel}, leading to the third line, 
where $\partial_\mu$ is the forward difference operator on the lattice.
\footnote{Another option about the choice of the 
axial vector current is to use the current induced from the 
symmetric part of the Dirac operator 
$\frac{1}{2} \left[ D, \gamma_5 \right] \gamma_5$. 
In this choice, as argued by P.~Hasenfratz\cite{no-tuning-mixing}, 
the explicit breaking term of the action do not  induce
the additive correction to quark mass, the renormalization of
axial and vector currents and the mixing between four-fermion 
operators with wrong chirality, except the axial 
anomaly\cite{index-finite-lattice}.}
In a practical point of view, however, 
it would be useful to have the explicit form of the axial vector
current, \footnote{The covariant current in the overlap formalism
has been discussed by Neuberger in
\cite{geometrical-aspect-of-anomaly}, 
and \cite{lattice-chirality-lattice98}. The relation between 
the covariant current and the Noether current obtained here 
will be discussed in a separate paper
\cite{an-exact-chiral-symmetry-from-overlap}. }
at least, in the case with the Neuberger-Dirac operator.
We will derive the explicit formula of the vector and axial vector 
currents in this case and also in the case with the Dirac 
operator in the rational approximation of 
Ref. \cite{practical-use-overlap-D}. 
Then, we discuss an Euclidean proof of the Nambu-Goldstone theorem of 
the spontaneous breakdown of the exact chiral symmetry,
using the axial Ward-Takahashi identity.

We first note from Eq.~(\ref{eq:axial-current-divergence}) 
that the axial vector current has a simple relation to
the vector current which is defined by 
\begin{eqnarray}
\label{eq:vector-current-divergence}
\delta S_F 
&=& a^4 \sum_{nml}
\bar \psi_n \left\{ D_{nl} \alpha_l - \alpha_n D_{nl} \right\} \psi_m 
\nonumber\\
&=& a^4 \sum_l \left\{ \partial_\mu \alpha_l  \, 
\sum_{nm} \bar \psi_n K_{l\mu}(n,m) \, \psi_m \right\}.
\end{eqnarray}
Once the vector current is constructed, 
the kernel of the axial vector current is obtained 
simply by the insertion of the operator
$\gamma_5 (1-aD)$:
\begin{equation}
\label{eq:vector-axial-relation}
  K^5_{l\mu}= K_{l\mu} \cdot \gamma_5 \left(1-aD\right) .
\end{equation}

The above relation can naturally be understood 
if we recall the fact that 
the operator $\gamma_5 \left(1-aD\right)$ satisfies 
the identity 
\begin{equation}
  \left\{ \gamma_5 \left(1-aD\right) \right\}^2 = 1 ,
\end{equation}
which follows from the Ginsparg-Wilson relation, 
as discussed by Narayanan \cite{GW-relation-to-factor-overlap} 
and its eigenvalues may be regarded to define another chirality 
analogous to $\gamma_5$, as considered by Hasenfratz, Niedermayer 
and L\"uscher \cite{chiral-decomposition}. This chirality
depends on the gauge fields. 
In fact, if we define the chiral components of $\psi$ and $\bar \psi$
through $\gamma_5 \left(1-aD\right)$ and $\gamma_5$, respectively,
\begin{eqnarray}
\label{eq:chiral-components-psi}
\gamma_5 \left(1-aD\right) \psi_\pm &=& \pm \psi_\pm, \\
\label{eq:chiral-components-bar-psi}
\bar \psi_{L,R} \, \gamma_5 &=& \pm \, \bar \psi_{L,R} .
\end{eqnarray}
the fermion action is decomposed into chiral parts as follows:
\begin{equation}
\label{eq:chiral-decomposition-action}
  \bar \psi D \psi 
= \bar \psi_{R} P_L D \psi_+ + \bar \psi_{L} P_R D \psi_- .
\end{equation}
This leads to the factorization of the partition function into
chiral determinants in the overlap 
formalism \cite{GW-relation-to-factor-overlap}.
In the kernel of the axial vector current 
of Eq.~(\ref{eq:vector-axial-relation}), 
the operator $\gamma_5 (1-aD)$ is to assign charges 
to $\psi_\pm$ opposite each other, contrary to the vector charges.

In this respect, we also notice the following fact.
For a chiral transformation of the type given by L\"uscher
to lead the exact symmetry of the action by the Ginsparg-Wilson 
relation, the weights of the Dirac operator in the transformation 
for $\psi$ and $\bar \psi$ should sum up to unity, 
but otherwise may be arbitrary. 
We can easily see by considering ``local'' transformations
following the Noether's procedure, that all such chiral transformations 
lead to the same axial vector current as given by
Eq.~(\ref{eq:axial-current-divergence}) and the associated axial 
Ward-Takahashi identities are physically equivalent.  
For example, the following chiral transformation 
also leaves the action invariant 
\begin{equation}
\label{eq:exact-chiral-symmetry-from-overlap}
\delta \psi_n = \sum_m \gamma_5 
    \left( 1 -  a D \right)_{nm} \psi_m , 
\quad
\delta \bar \psi_n = \bar \psi_n \gamma_5 ,
\end{equation}
and gives more natural definition of chiral transformation
in view of the chiral decomposition 
Eqs.~(\ref{eq:chiral-components-psi}) and 
(\ref{eq:chiral-components-bar-psi}), 
as discussed in \cite{chiral-decomposition}.
We will consider this version of the exact chiral symmetry 
later in our discussion of the spontaneous chiral symmetry breaking.

This particular chiral symmetry, in the case of the Neuberger-Dirac 
operator, can be regarded as a direct consequence of
the symmetry of quantum Hamiltonian in the underlying 
overlap formalism \cite{overlap}:
\begin{eqnarray}
\label{eq:overlap-hamiltonian-D}
{\cal H}_D &=& a^4 \sum_{nm}\hat a_n^\dagger  H_{nm} \hat a_m 
              -a^4 \sum_{nm}\hat b_n^\dagger  H_{nm} \hat b_m ,  \\
\label{eq:overlap-hamiltonian-5}
{\cal H}_{D5} &=& a^4 \sum_n \hat a_n^\dagger  \gamma_5 \hat a_n
                 -a^4 \sum_n \hat b_n^\dagger  \gamma_5 \hat b_n ,
\end{eqnarray}
This can be seen from the relation
between the Dirac field variables $\psi_n$ and $\bar \psi_n$ and 
the grassman representatives of the canonical variables
$(\hat a_n, \hat a_n^\dagger)$ and
$(\hat b_n, \hat b_n^\dagger)$ \cite{an-exact-chiral-symmetry-from-overlap}:
\begin{eqnarray}
\label{eq:dirac-variable-canonical-variable}
\psi_n &=& 
\frac{1}{\sqrt{2}}
\sum_m
\left\{
 \frac{1}{2} \left( 1+ \frac{H}{\sqrt{H^2}}\right)_{nm} a_m 
+\frac{1}{2} \left( 1- \frac{H}{\sqrt{H^2}}\right)_{nm} b_m 
\right\}
, \\
&& \nonumber\\
\bar \psi_n 
&=&  
\frac{1}{\sqrt{2}}\left\{
a_n^\dagger \left( \frac{1+\gamma_5}{2} \right) 
          + b_n^\dagger \left( \frac{1-\gamma_5}{2} \right) 
\right\} .
\end{eqnarray}
Then the symmetry transformation of the canonical variables 
\begin{equation}
\label{eq:symmetry-overlap-hamiltonian-axial}
  \delta \hat a_n = - \, \hat a_n, \quad 
  \delta \hat a_n^\dagger = \, \hat a_n^\dagger , \quad
  \delta \hat b_n =\, \hat b_n , \quad
  \delta \hat b_n^\dagger = -\, \hat b_n^\dagger ,
\end{equation}
in fact induces Eq.~(\ref{eq:exact-chiral-symmetry-from-overlap}).

Now let us construct the vector current in the case of the 
Neuberger-Dirac operator. The covariant transformation 
of the Dirac operator under the local $U(1)$ transformation
\begin{equation}
D_{nm} \longrightarrow 
D_{nm}^\prime = e^{ - i \alpha_n} \, D_{nm} \, e^{ i \alpha_m} 
\end{equation}
is induced by the local transformation of the overlap 
Hamiltonian $H$ 
\begin{equation}
H_{nm} \longrightarrow 
H_{nm}^\prime = e^{ - i \alpha_n} \,  H_{nm} \, e^{ i \alpha_m} 
\end{equation}
as 
\begin{equation}
a  D_{nm}^\prime= \left( 1 
+ \gamma_5 \frac{H^\prime}{\sqrt{\left(H^\prime\right)^2}} \right)_{nm} .
\end{equation}
Then, at the first order in the infinitesimal transformation parameter
$\alpha$, we have 
\begin{eqnarray}
\label{eq:vector-current-div-kernel}
a D_{nm} \, \alpha_m - \alpha_n \, a D_{nm}
&=& \gamma_5  \left\{ 
\left(\frac{H}{\sqrt{H^2}} \right)_{nm} \alpha_m
- \alpha_n \left(\frac{H}{\sqrt{H^2}} \right)_{nm} \right\} \nonumber\\
&=& \gamma_5 \, \delta \, \left(\frac{H}{\sqrt{H^2}} \right)_{nm} 
\biggl\vert
_{ \delta H = \{ H_{kl} \alpha_l - \alpha_k H_{kl} \} } , 
\end{eqnarray}
where 
\begin{eqnarray}
\delta H 
&=& H_{nm} \, \alpha_m - \alpha_n \, H_{nm}  
= \sum_l \partial_\mu \alpha_l  \, W_{l\mu}(n,m),  \\ 
&&\nonumber\\
W_{l\mu}(n,m)
&=& \gamma_5 \left\{ 
\frac{1}{2} \left(\gamma_\mu-1\right)
                 \delta_{nl}\delta_{l+\hat\mu, m} \, U_{l\mu}
+ \frac{1}{2} \left(\gamma_\mu+1\right)
 \delta_{n,l+\hat\mu}\delta_{lm} \, U_{l+\hat\mu,\mu}^\dagger 
\right\}. \nonumber\\
\end{eqnarray}
Using the integral representation,
\begin{equation}
\label{eq:integral-representation}
  \frac{1}{\sqrt{H^2}} 
= \int^\infty_{-\infty} \frac{dt}{\pi}   \frac{1}{t^2 + H^2} ,
\end{equation}
we obtain
\begin{eqnarray}
\delta \, \left(\frac{H}{\sqrt{H^2}} \right)_{nm} 
&=& \delta H \frac{1}{\sqrt{H^2}} 
- H \int^\infty_{-\infty} \frac{dt}{\pi}  
\frac{1}{t^2 + H^2} \left(H \delta H  + \delta H H \right)
\frac{1}{t^2 + H^2} \nonumber\\
&=& \int^\infty_{-\infty} \frac{dt}{\pi}  
\frac{1}{t^2 + H^2} \left(t^2 \delta H - H \delta H H \right)
\frac{1}{t^2 + H^2} .
\end{eqnarray}
Therefore, the kernel of the vector current is given by
\begin{equation}
\label{eq:vector-current-kernel}
a K_{l\mu}(n,m)
= \gamma_5 \left\{
\int_{-\infty}^{\infty} \frac{dt}{\pi}
\frac{1}{ (t^2+H^2)}
\left( t^2 W_{l\mu} - H W_{l\mu} H \right)
\frac{1}{ (t^2+H^2) } 
\right\}_{nm} . 
\end{equation}

It also follows from the Ginsparg-Wilson relation that 
the kernel of the vector current satisfies an identity given by
\begin{equation}
K_\mu \cdot \gamma_5 \left(1-aD\right) 
+ \left(1-aD\right) \gamma_5 \cdot K_\mu = 0. 
\end{equation}

Accordingly, using the relation Eq.~(\ref{eq:vector-axial-relation}), 
the axial vector current is obtained as
\begin{equation}
\label{eq:axial-current}
A_{l \mu} \equiv 
\sum_{nm} \bar \psi_n  K^5_{l\mu}(n,m) \psi_m ,
\end{equation}
\begin{equation}
\label{eq:axial-current-kernel}
K^5_{l\mu}(n,m)
= \left\{ 
K_{l\mu} \cdot
\left(-\frac{H}{\sqrt{H^2}} \right)\right\}_{nm} . 
\end{equation}
The flavor non-singlet currents can be obtained by 
including the generators $T^a$ of the rotation in the flavor space
in the above expression. 

Here we comment on the factorization property of the vector and 
axial currents into left- and right-handed currents. 
The vector and axial vector currents 
in Eqs. (\ref{eq:axial-current-divergence}) and 
(\ref{eq:vector-current-divergence})
are defined by introducing the local transformations 
\begin{eqnarray}
\delta \psi_n = \alpha_n \psi_n, \quad
\delta \bar{\psi}_n = - \alpha_n \bar{\psi}_n ,
\label{eqn:vt}
\end{eqnarray}
for vector current and 
\begin{eqnarray}
\delta \psi_n = \alpha_n \sum_m \gamma_5 (1-aD)_{nm} \psi_m , \quad
\delta \bar{\psi}_n = \alpha_n \bar{\psi}_n \gamma_5, 
\label{eqn:avc}
\end{eqnarray}
for axial vector current, respectively, 
and the kernels of these currents thus constructed 
respect the simple relation Eq.~(\ref{eq:vector-axial-relation}).
(Here we have adopted the chiral transformation
Eq. (\ref{eq:exact-chiral-symmetry-from-overlap}) rather than 
Eq. (\ref{eq:exact-chiral-symmetry}) for the reason explained before.)
In order to obtain the right-handed chiral current, for example, we may 
combine these two transformation as 
\begin{equation}
\delta \psi_n = \alpha_n \sum_m \left(P_+\right)_{nm} \psi_m , \quad
\delta \bar{\psi}_n = - \alpha_n \bar{\psi}_n P_L,
\end{equation}
where
\begin{equation}
  P_\pm= \frac{1\pm \gamma_5(1-aD)}{2}, 
\quad P_{R,L}=\frac{1\pm\gamma_5}{2} .
\end{equation}
Then we obtain the right-handed current as 
\begin{equation}
\label{eq:right-handed-current}
J^+_{l \mu} =
\frac{1}{2} \left( V_{l \mu} + A_{l \mu} \right)
= \sum_{nm} \bar \psi_n  \left(K_{l\mu} \cdot P_+ \right)_{nm} \psi_m .
\end{equation}
In the point of view of the chiral decomposition of Eqs. 
(\ref{eq:chiral-components-psi}), 
(\ref{eq:chiral-components-bar-psi}) 
and (\ref{eq:chiral-decomposition-action}), however, 
this current is not completely chiral since it contains 
not only the right-handed components $( \psi_+, \bar \psi_R)$ 
but also $\bar \psi_L$. 
This is because $P_{\pm}$ do not commute with the local variation 
operation (or operator $\delta$ )
and the above local transformation is not restricted to 
the right-handed chiral component $(\bar{\psi}_R,\psi_+) $ only.
To obtain the factorized chiral current which consists of only the
right-handed components $(\psi_+, \bar \psi_R)$, 
we may consider the right-handed local chiral transformation 
given as follows:
\begin{eqnarray}
\delta \psi_n =  \sum_{ml} \left( P_+ \right) _{nl} 
                \alpha_l  \left( P_+ \right)_{lm}  \psi_m , 
\quad
\delta \bar{\psi}_n = - \alpha_n \bar{\psi}_n P_L .
\label{eqn:lct}
\end{eqnarray}  
Here $P_+$ in front of the local parameter $\alpha_n$ is necessary 
so that the change of the variable along the chiral transformation
is consistent with the chiral decomposition:
\begin{equation}
  \psi_{n+}^\prime = \psi_{n+} + \delta \psi_{n} .
\end{equation}
Then we obtain the factorized right-handed current 
\begin{equation}
\label{eq:right-handed-current-overlap}
J^{R+}_{l \mu} = 
\sum_{nm} \bar \psi_n  
\left( P_L \cdot K_{l\mu} \cdot P_+ \right)_{nm} \psi_m .
\end{equation}
The transformation Eq.~(\ref{eqn:lct}) gives rise to half of the 
covariant anomaly as the Jacobian factor of the path integral measure. 
The local transformation for the other component 
$(\psi_- , \bar{\psi}_L)$ and its current are obtained in the same way. 
As it should be, the axial vector like combination of these local 
transformations is reduced to 
Eq. (\ref{eq:exact-chiral-symmetry-from-overlap}) when 
the transformation parameter is space independent. 
These currents are easily generalized to the case of the flavor 
non-singlet chiral transformations $SU(N_f)_R\times SU(N_f)_L$.

As we can see from
Eq.~(\ref{eq:dirac-variable-canonical-variable}), 
in the case of the Neuberger-Dirac operator, 
the local transformation 
Eq.~(\ref{eqn:lct}) can be regarded to be induced from 
the local transformation of the canonical variable 
$(\hat a_n, \hat a_n^\dagger)$ in the overlap formalism. 
Therefore the covariant and chiral (right-handed)
current Eq.~(\ref{eq:right-handed-current-overlap})
corresponds to the covariant current in the overlap formalism 
\cite{geometrical-aspect-of-anomaly,
lattice-chirality-lattice98,
an-exact-chiral-symmetry-from-overlap}.



Our expression of the axial vector current given by
Eqs.~(\ref{eq:vector-current-kernel}), (\ref{eq:axial-current}) 
and (\ref{eq:axial-current-kernel}) contains an integration over 
the parameter $t$. 
For analytical studies, for example, the weak coupling expansion, 
this integration can be explicitly 
performed \cite{weak-coupling-expansion-overlap-D}.
In numerical applications, however, we would need to treat it in some 
approximation: the integration may be replaced by a certain discrete 
summation.  Such an approach is closely related to the rational 
approximation of the Dirac operator 
considered by Neuberger 
\cite{practical-use-overlap-D}.\footnote{We would like to thank 
H.~Neuberger for suggesting this relation.} 
In fact, if we change the variable $t$ to the variable which has a 
compact region as
\begin{equation}
t = \tan \frac{\pi}{2} \sigma , \qquad \sigma \in [-1,1] ,
\end{equation}
the integral representation 
Eq.~(\ref{eq:integral-representation}), for example, may be rewritten as
\begin{equation}
\label{eq:inverse-square-root-integral-representation}
  \frac{1}{\sqrt{H^2}} 
= \int^\infty_{-\infty} \frac{dt}{\pi}   \frac{1}{t^2 + H^2} 
= \int_0^1 d \sigma 
  \frac{1}{\cos^2 \frac{\pi}{2}  \sigma }
  \frac{1}{\tan^2 \frac{\pi}{2}  \sigma + H^2 }.
\end{equation}
By discretizing the variable $\sigma$ as 
\begin{equation}
  \sigma \longrightarrow \frac{1}{n} \left(s-\frac{1}{2}\right) , 
\quad s=1,2,\ldots, n, 
\end{equation}
we are lead to the rational approximation to the Neuberger-Dirac
operator,
\begin{equation}
\label{eq:rational-approximation-D}
a  D^{(n)}
= 1 + \gamma_5 \left( \frac{H}{\sqrt{H^2}} \right)^{(n)} ,
\end{equation}
and 
\begin{equation}
\label{eq:rational-approximation-H}
\left(  \frac{H}{\sqrt{H^2}} \right)^{(n)}
= H \cdot \frac{1}{n} 
\sum_{s=1}^n \frac{1}{\cos^2 \frac{\pi}{2n}(s-\frac{1}{2})} 
\frac{1}{\tan^2 \frac{\pi}{2n}(s-\frac{1}{2}) + H^2} .
\end{equation}
Similarly, we may replace the integration over the
parameter $t$ in the expressions of 
the current kernels to the discrete summation. 
The same results are obtained using the expressions 
Eqs.~(\ref{eq:rational-approximation-D}) 
and (\ref{eq:rational-approximation-H}) from the beginning.
Thus we obtain the axial vector current in the case 
with the Neuberger-Dirac operator in the rational approximation 
as follows:
\begin{eqnarray}
\label{eq:vector-current-kernel-rational}
a K_{l\mu}^{(n)}
&=&\gamma_5 
\frac{1}{n} 
\sum_{s=1}^n \frac{1}{\cos^2 \frac{\pi}{2n}(s-\frac{1}{2})}
\, \frac{1}{\tan^2 \frac{\pi}{2n}(s-\frac{1}{2})+H^2}
\times
\nonumber\\
&& \qquad\qquad
\left(\tan^2 \frac{\pi}{2n}(s-\frac{1}{2}) \, W_{l\mu}
   -  H W_{l\mu} H \right)
\frac{1}{\tan^2 \frac{\pi}{2n}(s-\frac{1}{2})+H^2} .
\nonumber\\
&&\\
\label{eq:axial-current-kernel-rational}
K_{l\mu}^{5 (n)}
&=& K_{l\mu}^{(n)} \left(-\frac{H}{\sqrt{H^2}} \right)^{(n)} .
\end{eqnarray}

It may also be possible to use the expansion in Legendre polynomials 
for the inverse square root of $H^2$ as adopted by Hernandes, Jansen 
and L\"uscher \cite{ledendre-polynomial-expansion,
locality-overlap-D-small-su3}. 
With the help of the integral representation 
Eq.~(\ref{eq:inverse-square-root-integral-representation}), 
we can derive the expansion of the current kernels in terms of
Legendre polynomials. 

Finally, we consider the axial Ward-Takahashi identity
associated with the flavor non-singlet chiral transformation
Eq.~(\ref{eq:exact-chiral-symmetry-from-overlap}), 
\begin{equation}
\delta \psi_n = \sum_m T^a \gamma_5 
    \left( 1 -  a D \right)_{nm} \psi_m , 
\quad
\delta \bar \psi_n = \bar \psi_n \gamma_5 T^a . 
\end{equation}
In particular, we discuss the chiral multiplet of 
the scalar and pseudo-scalar density operators 
considered by Neuberger, Hasenfratz and Chandrasekharan 
\cite{almost-massless,no-tuning-mixing,soft-pion-theorem-from-GW},
\begin{equation}
\delta \, \left( \bar \psi_m \gamma_5 \psi_m \right) = 
2 \, \, \bar \psi_m \sum_l \left( 1-\frac{a}{2} D \right)_{ml} \psi_l .
\end{equation}
Its axial Ward-Takahashi identity is given as follows:
\begin{equation}
\label{eq:chiral-WT-id-with-pseudo-scalar}
a^4 \, \partial_\mu^\ast \left\langle A_{n\mu}^a \, \,
\bar \psi_m T^b \gamma_5 \psi_m 
\right\rangle
+\delta_{nm}  \delta^{ab}
\left\langle 
\bar \psi_m \sum_l \left( 1-\frac{a}{2} D \right)_{ml} \psi_l 
\right\rangle = 0 .
\end{equation}
Here we assume that the vacuum expectation values are defined 
by the functional integrals over both the fermion and 
the gauge fields. In the case of the Neuberger-Dirac operator, 
we also assume that the gauge fields satisfies the smoothness 
condition discussed in \cite{locality-overlap-D-small-su3} by the
choice of an appropriate gauge field action. 
Note that, with this constraint, the kernel of the axial vector 
current can be shown to be exponentially (uniformly) bounded 
\cite{locality-overlap-D-small-su3}.

Then the identity implies that the massless pole would appear in 
the correlation function of the axial vector current 
and the pseudo scalar density 
($\hat p_\mu \equiv \frac{2}{a} \sin \frac{a p_\mu}{2}$) 
\begin{eqnarray}
{\rm F.T.} \, \left\langle A_{n\mu}^a \, \,  
\bar \psi_m T^b \gamma_5 \psi_m 
\right\rangle ( p ) 
&=& 
-i \, \frac{ \hat p_\mu }{\hat p^2} \, 
\, \delta^{ab} \left\langle 
\bar \psi_m \sum_l \left( 1-\frac{a}{2} D \right)_{ml} \psi_l 
\right\rangle  ,
\nonumber\\
\end{eqnarray}
if the spontaneous breakdown of the exact chiral symmetry occurs
\begin{equation}
\left\langle \bar \psi_m 
  \sum_l  \left( 1-\frac{a}{2} D \right)_{ml} \psi_l \right\rangle
\not = 0.  
\end{equation}
This consists an Euclidean proof of the Nambu-Goldstone theorem 
on the lattice. 
Let us assume the coupling of the Nambu-Goldstone 
boson to the axial vector current and the pseudo scalar density as follows:
\begin{eqnarray}
\langle \, 0 \, \vert \, 
A_{n\mu}^a 
\,  \vert \, \pi^b(p) \, \rangle
&=& \delta^{ab} \,  F_\pi \, i \hat p_\mu  \, e^{ipn} , \\ 
\langle \, 0 \, \vert \, 
\bar \psi_n T^a \gamma_5 \psi_n 
\,  \vert \, \pi^b(p) \, \rangle
&=& \delta^{ab} \, a^{-2} \, Z_\pi^{\frac{1}{2}} \, e^{ipn} .
\end{eqnarray}
Then we obtain the relation,
\begin{equation}
\label{eq:relation-fpi-condensate}
F_\pi Z_\pi^{\frac{1}{2}} = - \, a^2 \left\langle 
\bar \psi_m \sum_l \left( 1-\frac{a}{2} D \right)_{ml} \psi_l 
\right\rangle ,
\end{equation}
which is well-known in the continuum theory.

Comments about the choice of the multiplet
of the scalar and pseudo scalar density operators are in order.
Under the ``global'' chiral transformation 
Eq.~(\ref{eq:exact-chiral-symmetry-from-overlap}), 
it is easily verified that the following scalar and pseudo scalar 
density operators consist an exact chiral multiplet 
\cite{chiral-decomposition}:
\begin{equation}
S_n = \bar \psi_n  \sum_m \left( 1 - \frac{a}{2}D \right)_{nm} \psi_m ,
\quad
P_n = \bar \psi_n  \sum_m \gamma_5 \left( 1 - \frac{a}{2}D \right)_{nm}
\psi_m ,
\end{equation}
as
\begin{equation}
\delta  S_n  = 2 P_n ,  \qquad \delta  P_n  = 2 S_n .
\end{equation}
Then one might think of the axial Ward-Takahashi identity 
which involves the pseudo scalar density $P_n$. 
The above chiral property, however, is not reflected directly
in the axial Ward-Takahashi identity, because it is derived 
through the ``local'' chiral transformation. In fact, 
what we obtain in this case is as follows:
\begin{equation}
a^4 \,\partial_\mu^\ast \left\langle A_{n\mu} \, 
  P_m 
\right\rangle
+\delta_{nm} \left\langle 
S_m
\right\rangle 
+\left\langle 
\bar \psi_m 
\gamma_5 \left( 1-\frac{a}{2} D \right)_{mn} \, 
\sum_l \gamma_5 \left( 1-\frac{a}{2} D \right)_{nl} \psi_l
\right\rangle 
= 0 .
\end{equation}
The third term in the l.h.s. is not proportional to $\delta_{nm}$
times the chiral condensate, 
contrary to Eq.~(\ref{eq:chiral-WT-id-with-pseudo-scalar}). 
As far as the axial Ward-Takahashi identity for the multiplet of 
the scalar and pseudo scalar densities is concern,
the choice of Chandrasekharan gives rise to the simple identity.
%

In summary, we have considered the exact chiral symmetry and its 
spontaneous breakdown in lattice QCD 
with the Dirac operators satisfying the Ginsparg-Wilson relation. 
We have constructed explicitly the axial vector current
in the cases with the Neuberger-Dirac operator and its rational 
approximation, which turns out to be related to the vector current 
simply by the insertion of the operator $\gamma_5 (1-aD)$.
Then we have considered an Euclidean proof of the Nambu-Goldstone theorem
with the axial Ward-Takahashi identity associated with the ``local'' 
chiral transformation. The relation given by 
Eq.~(\ref{eq:relation-fpi-condensate}) or the explicit formula 
of the axial vector current may be useful in the numerical 
evaluation of the pion decay constant $F_\pi$. 

\section*{Acknowledgments}
We are indebted to H.~Neuberger 
for enlightening discussions.  
Y.K. would like to thank  M.~L\"uscher 
for the communication concerning the locality of Neuberger's 
Dirac operator. Y.K. is supported in part by Grant-in-Aid 
for Scientific Research from Ministry of Education, Science 
and Culture(\#10740116,\#10140214).
A.Y. would like to thank Pamela Morehouse for sending him an 
useful preprint.

\end{document}